\documentclass[manuscript]{aastex}
\usepackage{graphicx}

\shorttitle{Magnetized jets driven by the sun}
\shortauthors{Opher, Drake, Zieger and Gombosi}

\begin{document}

\title{Magnetized jets driven by the sun: the structure of the heliosphere revisited}

\author{M. Opher\altaffilmark{1,3}}
\affil{Astronomy Department, Boston University, Boston, MA 02215}
\email{mopher@bu.edu}

\author{J. F. Drake\altaffilmark{2}}
\affil{Department of Physics and the Institute for Physical Science and Technology, University of Maryland, College Park, MD} 
 
\author{B. Zieger\altaffilmark{3}}
\affil{Center for Space Physics, Boston University, Boston, MA 02215}

\author{T. I. Gombosi\altaffilmark{4}}
\affil{University of Michigan, Ann Arbor, MI}

\begin{abstract}
The classic accepted view of the heliosphere is a quiescent, comet-like shape aligned in the direction of the Sun's travel through the interstellar medium (ISM) extending for 1000’s of AUs (AU: astronomical unit). 
Here we show, based on magnetohydrodynamic (MHD) simulations, that the tension (hoop) force of the twisted magnetic field of the sun confines the solar wind plasma beyond the termination shock and drives jets to the North and South very much like astrophysical jets. These jets are deflected into the tail region by the motion of the Sun through the ISM similar to bent galactic jets moving through the intergalactic medium. The interstellar wind blows the two jets into the tail but is not strong enough to force the lobes into a single comet-like tail, as happens to some astrophysical jets (Morsony et al. 2013). Instead, the interstellar wind flows around the heliosphere and into equatorial region between the two jets. As in some astrophysical jets that are kink unstable (Porth et al. 2014) we show here that the heliospheric jets are turbulent (due to large-scale MHD instabilities and reconnection) and strongly mix the solar wind with the ISM beyond $400~AU$. The resulting turbulence has important implications for particle acceleration in the heliosphere. The two-lobe structure is consistent with the energetic neutral atoms (ENAs) images of the heliotail from IBEX (McComas et al. 2013) where two lobes are visible in the North and South and the suggestion from the CASSINI (Krimigis et al. 2009, Dialynas et al. 2013) ENAs that the heliosphere is ``tailless''.

\end{abstract}


\section{Introduction}
The long time view of the shape of the heliosphere is that it is a comet-like object (Parker 1961, Baranov \& Malama 1993) with a long tail opposite to the direction with which the solar system moves through the local interstellar medium. The solar magnetic field at large distance from the Sun is azimuthal forming a spiral (the so called  ``the Parker spiral'') as a result of the rotation of the Sun. The traditional picture of the heliosphere as a cometary-like structure comes from the assumption that even though the solar wind becomes subsonic at the termination shock as it flows down the tail it is able to stretch the solar magnetic field. 

Current magnetohydrodynamic (MHD) models (Opher et al. 2009, Alexashov et al. 2004, Ratkiewicz \& Webb 2002, Pogorelov et al. 2007, Washimi et al. 2011, Provornikova et al. 2014, Pogorelov et al. 2013) typically focus on benchmarking their results with the in-situ Voyager 1 and 2 results (Stone et al. 2005, 2008) as well with the energetic neutral atoms global maps by IBEX (McComas et al. 2009) and CASSINI (Krimigis et al. 2009). These observations led the modelers to focus their attention on the nose region of the heliosphere but their simulations also suggest that the tail is a comet-like, extending to long distances (Pogorelov et al. 2014). 

Here we show, based on MHD simulations, that the twisted magnetic field of the sun confines the solar wind plasma and drives jets to the North and South very much like some astrophysical jets. Astrophysical jets around massive black holes are thought to orginate from Keplerian accretion disks and are driven by centrifugal forces (Blandford \& Payne 1982). However, the jets in the case of the heliosphere are driven downstream of the termination shock similar to what was proposed for the Crab Nebula (Chevalier \& Luo 1994; Lyubarsky 2002). In this region of subsonic flow the magnetic tension (hoop) force is strong enough to collimate the wind. The tension force is also the primary driver of the outflow. The heliosheath is in force balance with the pressure (magnetic and particle) of the interstellar medium along the radial direction. In this direction the pressure gradient within the heliosheath is balanced by magnetic curvature. However, along the axis of the jet there is no curvature force so the resulting axial pressure gradient drives the axial flow of the jet. The heliospheric jets are deflected into the tail region by the motion of the Sun through the ISM similar to bent galactic jets moving through the intergalactic medium. The interstellar wind blows the two jets into the tail but is not strong enough to force the lobes into a single comet-like tail, as happens to some astrophysical jets (Morsony et al. 2013). Instead, the interstellar wind flows around the heliosphere and into equatorial region between the two jets. We show here that like some astrophysical jets (Porth et al. 2014) the heliospheric jets are turbulent.

The organization of this letter is as follows: we first discuss the model used, then the structure and formation of the {\it two-lobes} and finally the implications of the new picture of the heliosphere.

\section{Model}

Our model 5-fluid code is based on the 3D multi-fluid MHD code BATS-R-US with adaptive mesh refinement (Toth et al. 2012). It evolves 1 ionized and 4 neutral species of H as well as the magnetic field of the sun and the interstellar medium. We used a monopole configuration for the solar magnetic field. The multi-fluid approach for the neutrals captures the main features of the kinetic model (Izmodenov et al. 2009). Atoms of interstellar origin represent population 4. Population 1 appears in the region behind the bow shock (or slow shock, depending on the intensity of $B_{ISM}$(Zieger et al. 2013). Populations 3 and 2 appear in the supersonic solar wind and in the compressed region behind the Termination Shock, respectively. All four populations are described by separate systems of the Euler equations with corresponding source terms describing neutral-ion charge exchange.

The inner boundary of our domain is a sphere at $30~AU$ and the outer boundary is at $x = \pm1500~AU$, $y = \pm 1500~AU$, $z = \pm 1500~AU$. Parameters of the solar wind at the inner boundary at $30~AU$ are: $v_{SW} = 417 km/s$, $n_{SW} = 8.74 \times 10^{-3} cm^{-3}$, $T_{SW} = 1.087 \times 10^{5}~K$ (OMNI solar data, http://omniweb.gsfc.nasa.gov/). The Mach number of the solar wind is $7.5$ and is therefore super-fast-magnetosonic. Therefore all the flow parameters can be specified at this boundary. The magnetic field is given by the Parker spiral magnetic field (Parker 1958), 
\begin{equation}
B=B_{0} {\left ( \frac{R_{0}}{r} \right ) }^{2} e_{r} - B_{0} \left ( \frac{R_{0}^{2}}{r} \right ) \frac{\Omega sin \Theta}{v_{SW}} e_{\phi} 
\end{equation}
where $R_{0}$ is the inner boundary $30~AU$, $v_{SW}$ is the solar wind speed 
with the radial component $B_{SW} = 7.17 \times 10^{-3} nT$ at the equator at $30~AU$, $\Theta$ is the polar angle of the field line, and $\Omega$ is the equatorial angular velocity of the Sun. We assume that the magnetic axis is aligned with the solar rotation axis. 

The solar wind flow at the inner boundary is assumed to be spherically symmetric. For the interstellar plasma we assume: $v_{ISM} = 26.4~km/s$, $n_{ISM} = 0.06~cm^{-3}$, $T_{ISM}= 6519~K$. The number density of H atoms in the interstellar medium is $n_{H} = 0.18~cm^{-3}$, the velocity and temperature are the same as for the interstellar plasma. The coordinate system is such that the z-axis is parallel to the solar rotation axis, the x-axis is $5^{\circ}$ above the direction of interstellar flow with y completing the right-handed coordinate system. The grid was made up of $6.05 \times 10^{7}$ cells ranging in size from $0.37~AU$ at the inner boundary to $93.75~AU$ at the outer boundary. The tail region in the heliosheath had a resolution of $0.7~AU$ all the way to $x=1000~AU$ in the deep tail. The case with $B_{ISM}$ was run to $480,000$ time steps, which corresponds to 659 years. The case with no $B_{ISM}$ was run to $660,000$ time steps, which corresponds to 865 years.

The strength of the $B_{ISM}$ in the model is $4.4~\mu G$. The orientation of $B_{ISM}$ continues to be debated in the literature. The orientation of $B_{ISM}$ is defined by two angles, $\alpha_{BV}$ and $\beta_{BV}$. $\alpha_{BV}$ is the angle between the interstellar magnetic field and flow velocity of the interstellar wind and $\beta_{BV}$ is the angle between the $B_{ISM}$ - $v_{ISM}$ plane and the solar heliographic equator. To account for the heliospheric asymmetries, such as the different crossing distances of the termination shock by V1 and 2, a small value of $\alpha_{BV} \sim 10-20^{\circ}$ is required (Opher et al. 2009, Izmodenov et al. 2009). Others studies (Heerikhuisen \& Pogorelov 2011, Chalov et al. 2010) have used the observed shape and location of the IBEX ribbon to constrain the magnitude and orientation of $B_{ISM}$. However, such constraints are sensitive to the specific model of the IBEX ribbon, which continues to be uncertain. In any case for the present study the exact direction of $B_{ISM}$ and its intensity are not important. 

\section{Solar Magnetized Jets}

We performed 3-D MHD simulations showing that the heliosphere does not have a comet-like structure. The solar magnetic field was chosen to be unipolar (Opher \& Drake 2013) to avoid artificial numerical magnetic reconnection at the nose as well as in the solar equator across the heliospheric current sheet. We also present a simulation with an interstellar wind but with no interstellar magnetic field to avoid artificial reconnection at the heliopause interface. 

Even with no interstellar magnetic field the heliosphere develops a two-lobe structure organized by the solar magnetic field (Figure 1a-c). The lobes survive due to the resistance of the solar magnetic field to being stretched. The magnetic tension force must therefore be sufficiently strong to collimate the jets. To show this we estimate the tension on a field line with a radius of curvature $R$ as $F_{tension} = {|\bf B}\cdot \nabla {\bf B}|/4\pi \sim (B^{2}/8\pi)(2/R)$. So $F_{tension} \sim 2 P_{B}/R$, where $P_{B}$ is the magnetic pressure. The force stretching the magnetic field due to the flows is $F_{streatching} \sim \rho |{\bf v} \cdot {\bf \nabla v}|/2 \sim \rho v^{2} \kappa_v/2 \sim \rho v^{2}/2R \sim P_{ram}/R$ where $\kappa_v$ is like the curvature with $\kappa \sim 1/R$ and $P_{ram}$ is the ram pressure. So the ratio between the two forces $F_{streatching} /F_{tension} \sim P_{ram}/2P_{B}$, which is $< 1$ down the tail past the termination shock (Figure 1d). Thus, the magnetic tension (hoop stress) is sufficient to resist the streatching by the flows and can collimate jets. The result is a tail divided in two separate plasmas confined by the solar magnetic field (Figure 1a; 1c). The two lobes are separated by the pressure of the interstellar plasma that flows around the heliosphere and into the equatorial region downstream of the heliosphere (Figure 1a).  This behavior can be seen in Figure 1f where the meridional flows $U_y$ are shown and the ISM streamlines flow between the two lobes in Figure 1a. Thus, the interstellar wind is not sufficiently strong to force the North and South lobes of the heliosphere to merge together to form a comet-like structure. The thermal pressure from the ISM balances the magnetic and plasma pressure in the lobes in the y-z plane in the down-tail region.

In the heliosheath the plasma pressure is generally much higher than the magnetic pressure so it might seem surprising that the magnetic field controls the formation and structure of the jets. There are two factors that explain why the magnetic field and specifically the tension forces are critically important. First, due to the expansion of the plasma as it flows from the termination shock out towards the heliopause, the plasma pressure drops until the two pressures are comparable. Figure 3a shows the ratio between the two pressures in a cut in the meridional plane ($y=0$). Immediately after the termination shock the gas (thermal) pressure dominates (by almost an order of magnitude) but further out it becomes weaker due to expansion. Thus, the ratio of the magnetic to thermal pressure increases. Near the heliopause the system approaches approximate equipartition. On the other hand equipartition is not a requirement at the heliopause boundary. The ratio between the magnetic to thermal pressure at the heliopause depends on the value of the interstellar pressure compared with the thermal pressure downstream of the termination shock (see for example the plasma and magnetic profiles for the Crab in Fig. 1 of Begelman \& Li 1992). Second, even in a high-$\beta$ heliosheath it is the magnetic tension force that controls the total pressure drop from the termination shock to the heliopause. This was also noted in calculations related to the Crab Nebula (Begelman \& Li 1992). Since there is no tension force along the axis this same axial pressure drop is balanced by the inertia associated with the generation of the axial flow. This can be shown in a rigorous analytic calculation of the structure of the heliosheath and associated flow (Drake et al 2015). Similar forces have been proposed to drive the flows in the Crab Nebula, downstream of the termination shock (Chevalier \& Luo 1994) 

\begin{figure}[htbp]
\centering
\includegraphics[width=0.3\textwidth]{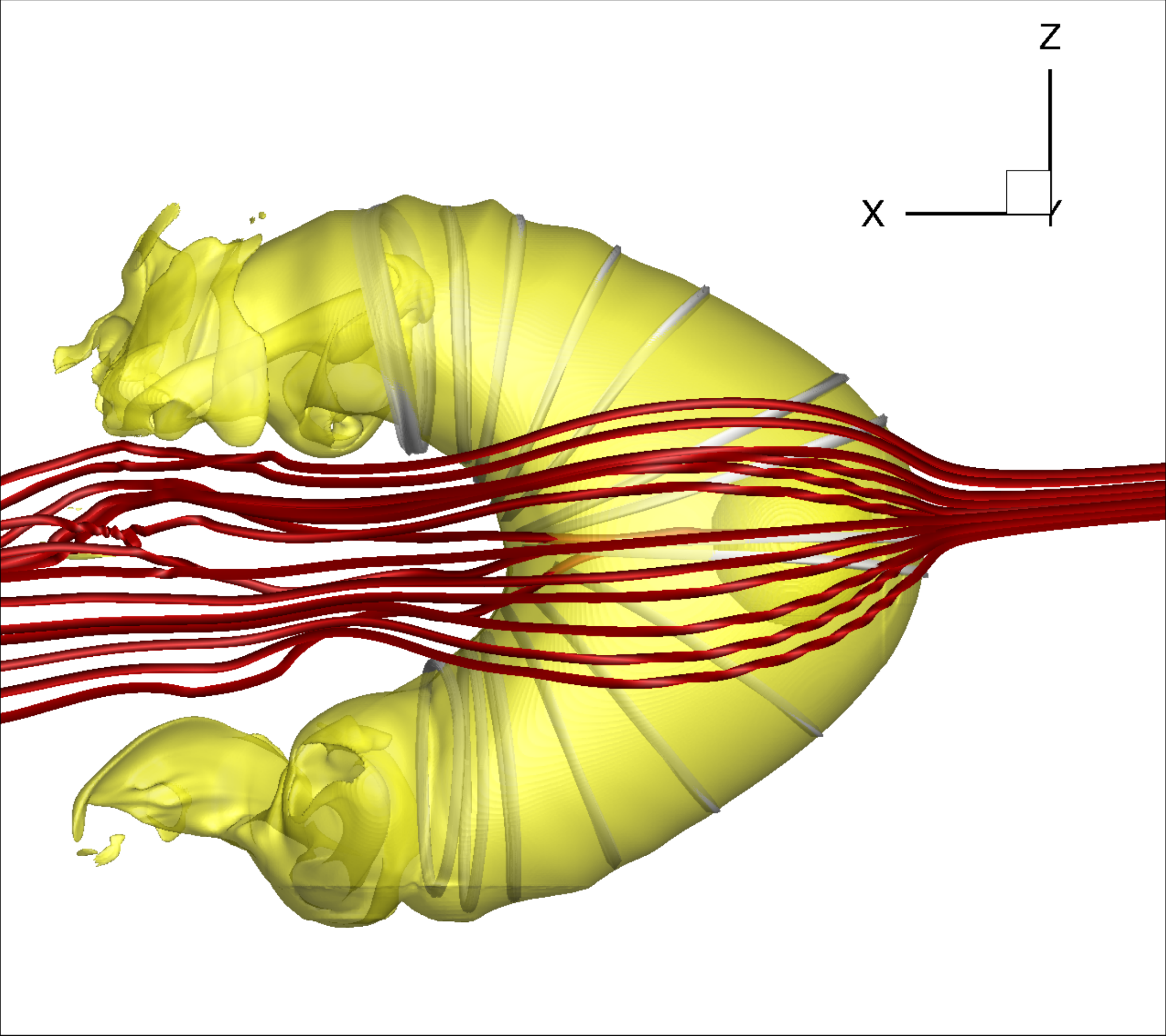}          
\includegraphics[width=0.3\textwidth]{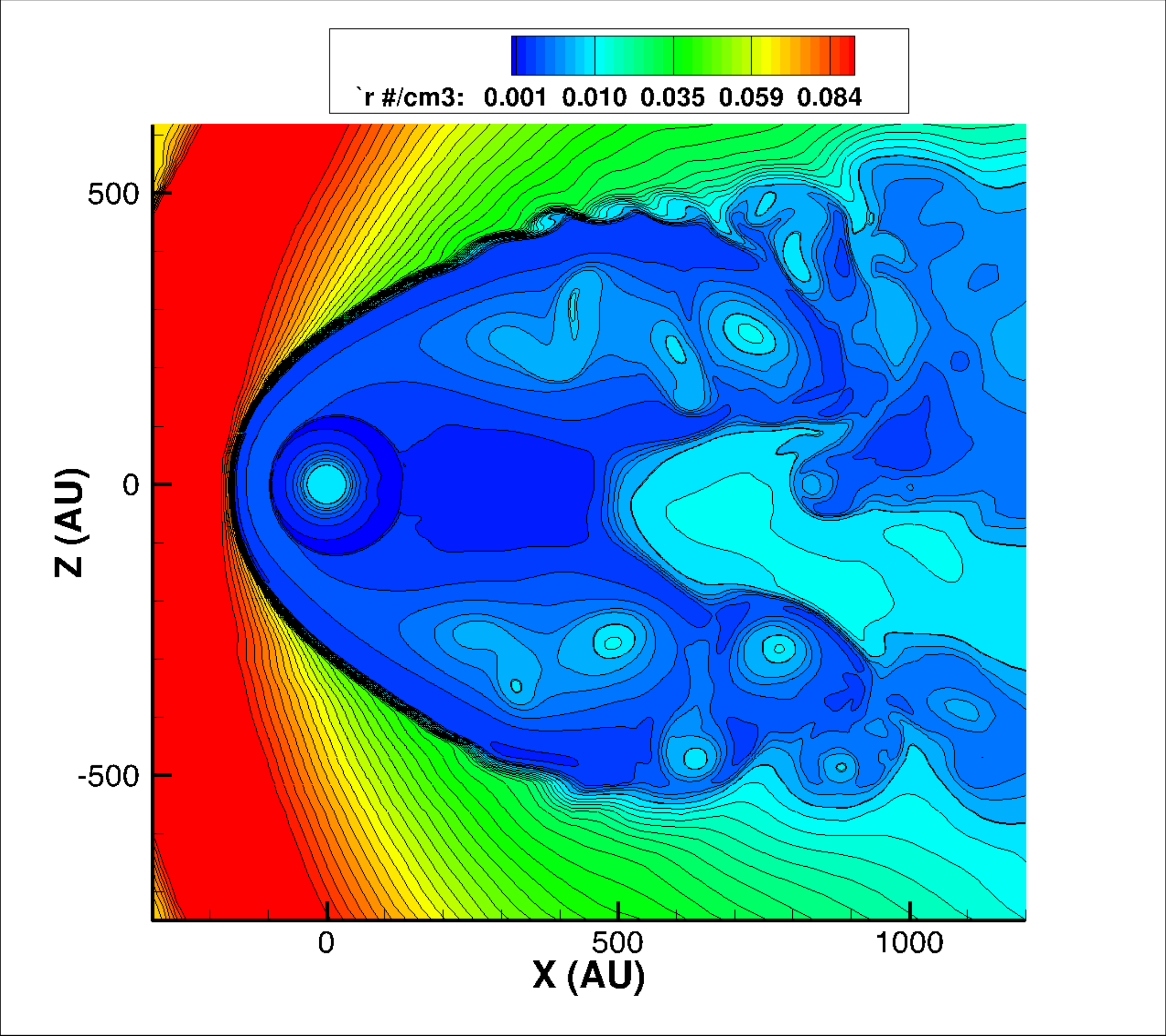}  
\includegraphics[width=0.3\textwidth]{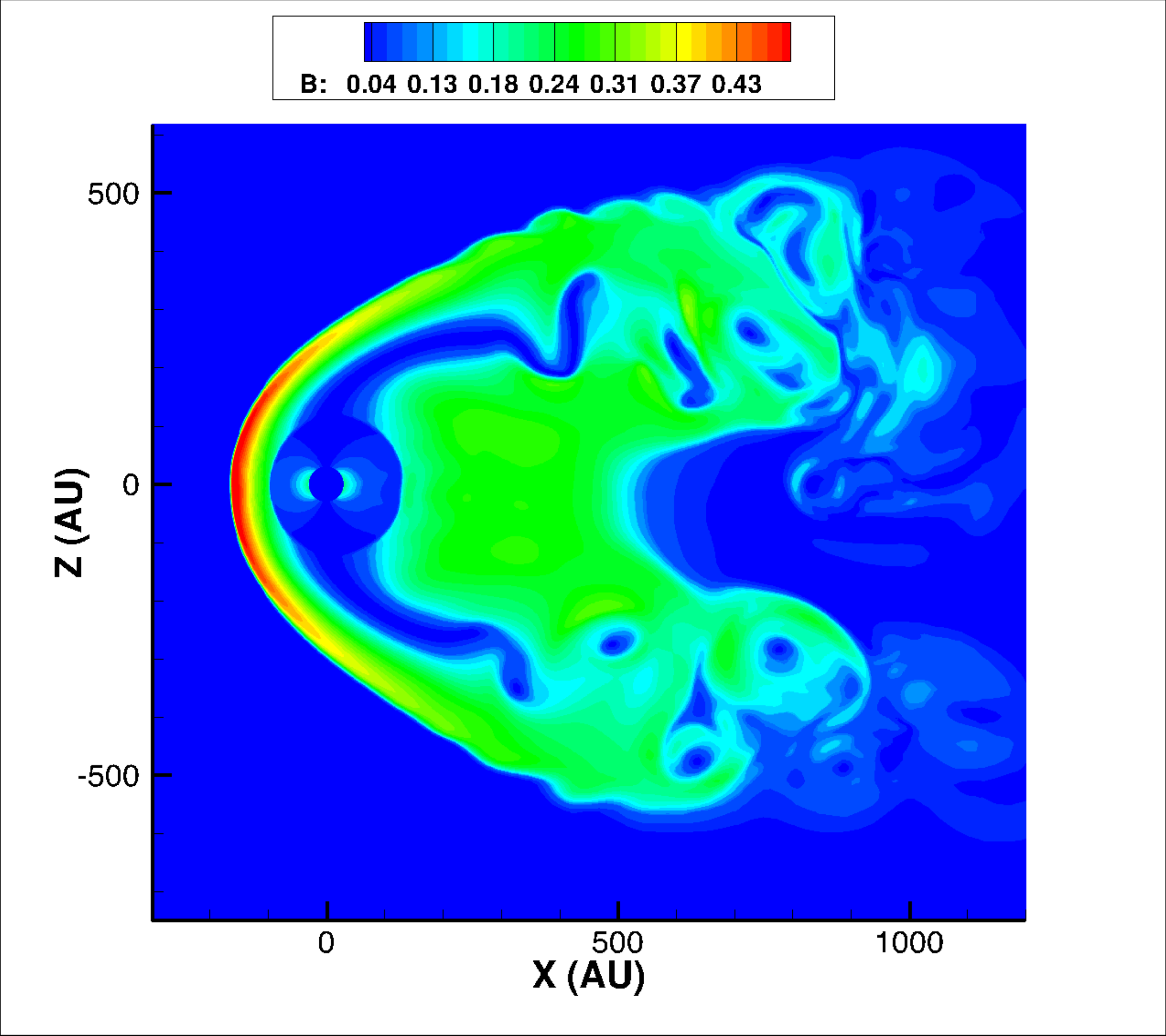}  
\includegraphics[width=0.3\textwidth]{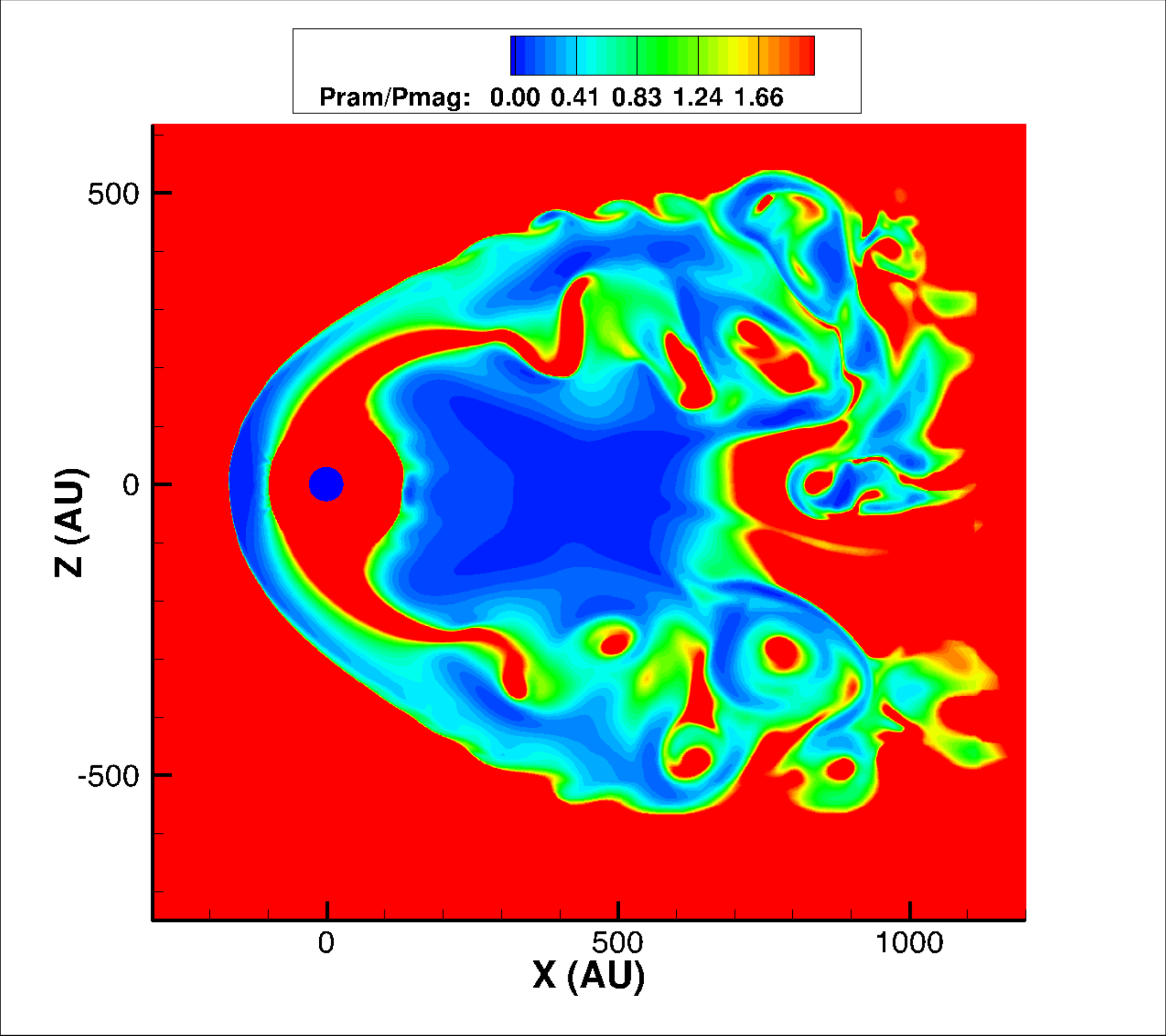}  
\includegraphics[width=0.3\textwidth]{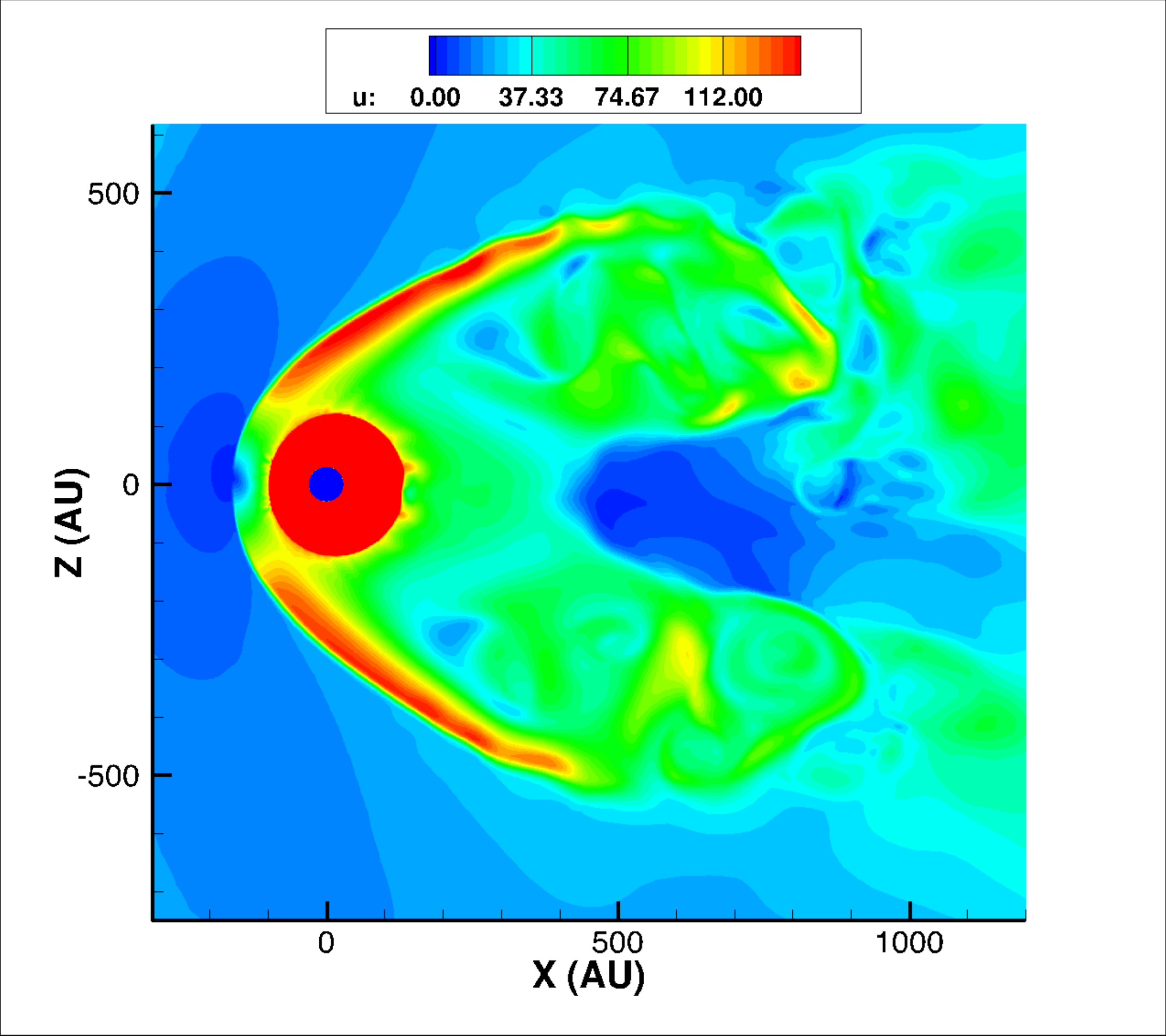}  
\includegraphics[width=0.3\textwidth]{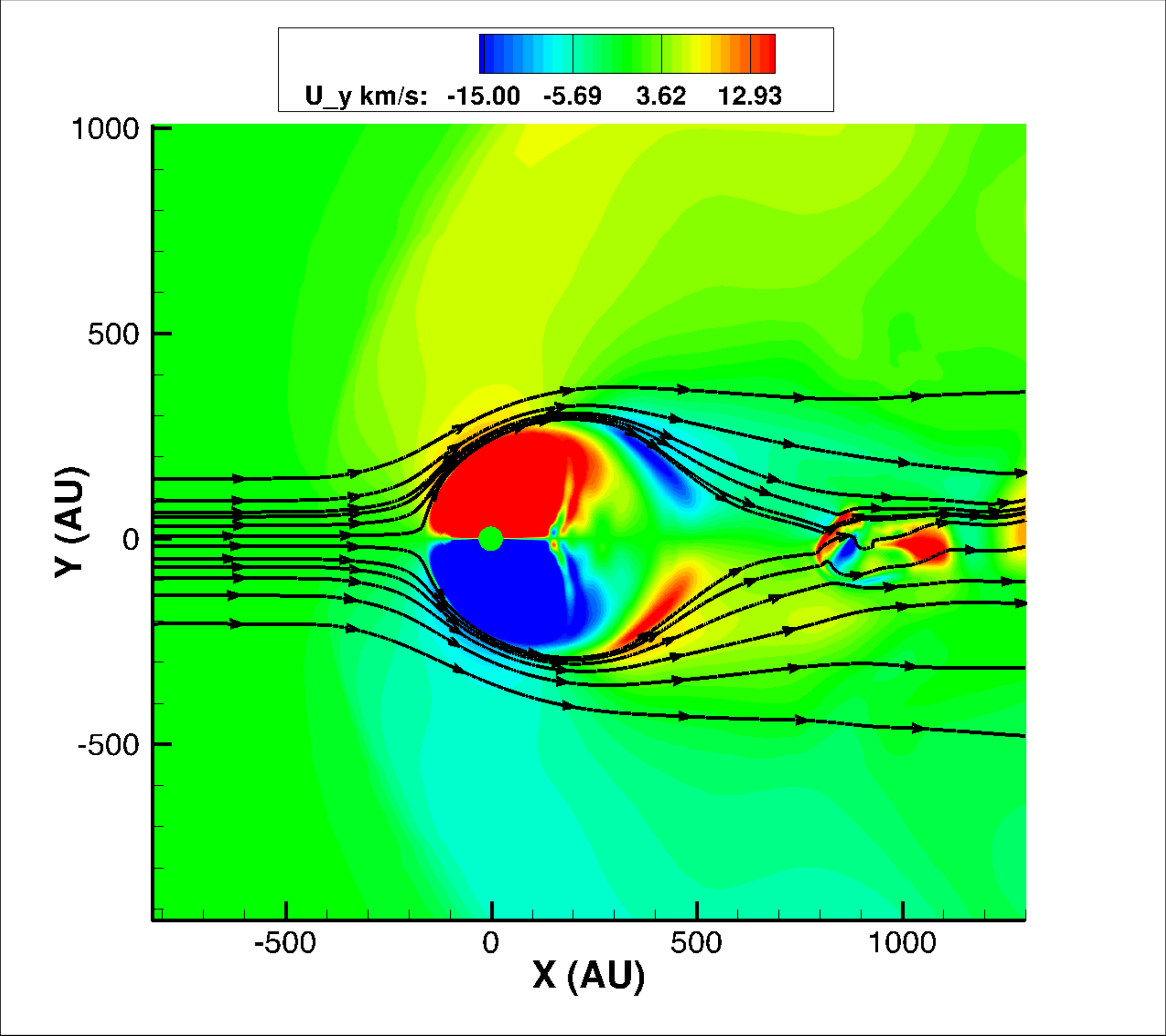}     
\caption{Two lobe structure - case with no interstellar magnetic field. The panels are shown at the end of the simulation at 865 years (a) View from the West from the point of view of looking from the ISM towards the nose of the heliosphere. The grey lines are the magnetic field lines and the red the interstellar wind velocity streamlines that streams from right to left in panel (a). The yellow iso-surface at $ln T=12.7$ denotes the heliopause. The panels (b)-(e) are in the meridional plane, at $y=0~AU$. Contours are: (b) density; (c) magnetic field; (d) $P_{ram}/P_{B}$; (e) speed; (f) cut in the equatorial plane, at $z=0 AU$; contours are the meridional flows $U_y$. The black lines are the velocity streamlines. The view from panel (a) is reversed from panels (b)-(e). }
\label{figure1}
\end{figure}

Further, there is evidence from the simulation data that the magnetic tension drives flows in the heliosheath. The solar magnetic field twists so that the azimuthal Parker field in the solar wind lies in the y-z plane in the downstream region. The tension in the solar magnetic field contracts and accelerates the heliosheath plasma downstream (Figure 1e $-$ the red contours at high latitudes). The direction of the strongest flow is in the North-South direction. 

The two lobes are unstable  - several instabilities are probably taking place. One is the Kelvin-Helmholtz (Wang \& Belcher 1998) instability since there is no interstellar magnetic field in this case to stabilize the instability. The two lobes are also prone to kink and sausage instabilities since the axial magnetic field in the lobes is much smaller than the azimuthal field, which is the driver for these instabilities. The instabilities drive flows so that the lobes become highly turbulent, which erodes the lobes as they flow tailward.

Porth \& Komissarov (2014) argue that the high expansion rate of astrophysical jets leads to a causal disconnection of the opposite sides of the jet and therefore might explain for some cases the absence of instabilities in these systems. The absence of instabilities in the solar wind upstream the termination shock in could be for the same reason. The flow is supersonic so large regions of the solar wind are causally disconnected. In the jets down stream of the termination shock the flows are sub-fast-magnetosonic so the jets are causally connected at their largest spatial scales. 

The thermal pressure from the ISM in the tail is the key factor that prevents the merger of the two lobes. We also completed a simulation with the same grid and conditions as in Fig. 1 but including the interstellar magnetic field. We chose the magnitude and orientation of the interstellar magnetic field from prior studies (Opher et al. 2009) that match the measured heliospheric asymmetries and with a direction that eliminates reconnection at the nose (Opher \& Drake 2013). This configuration reproduces the twist of the interstellar magnetic field towards the solar direction ahead of the heliopause as seen by the recent observations of Voyager 1 (Burlaga \& Ness 2014). 

The two lobes are present as well in this case (Figure 2) and again are organized by the solar magnetic field {\bf and its magnetic tension}. However, in this case the distance between the sun and the heliopause down the tail between the two lobes is much smaller ($250~AU$ as opposed to $560~AU$ in the case with no $B_{ISM}$). The lobes are more eroded as well as a result of instabilities and reconnection in the flanks (Figure 3). Again nearly everywhere downstream of the termination shock $P_{ram}/P_{B}  < 1$ so the solar magnetic field is strong enough to confine the flow.

\begin{figure}[htbp]
\centering
\includegraphics[width=0.45\textwidth]{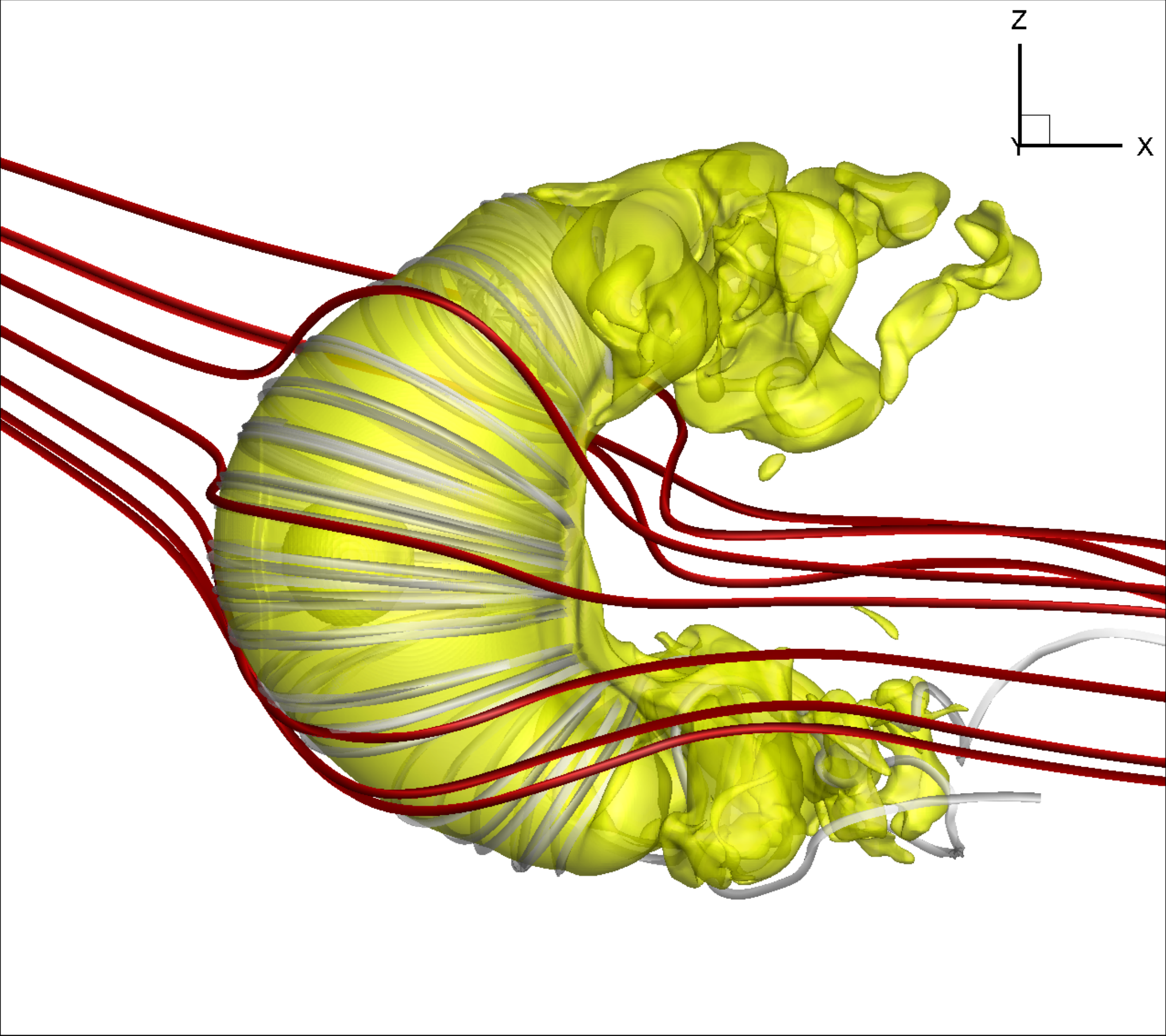}          
\includegraphics[width=0.45\textwidth]{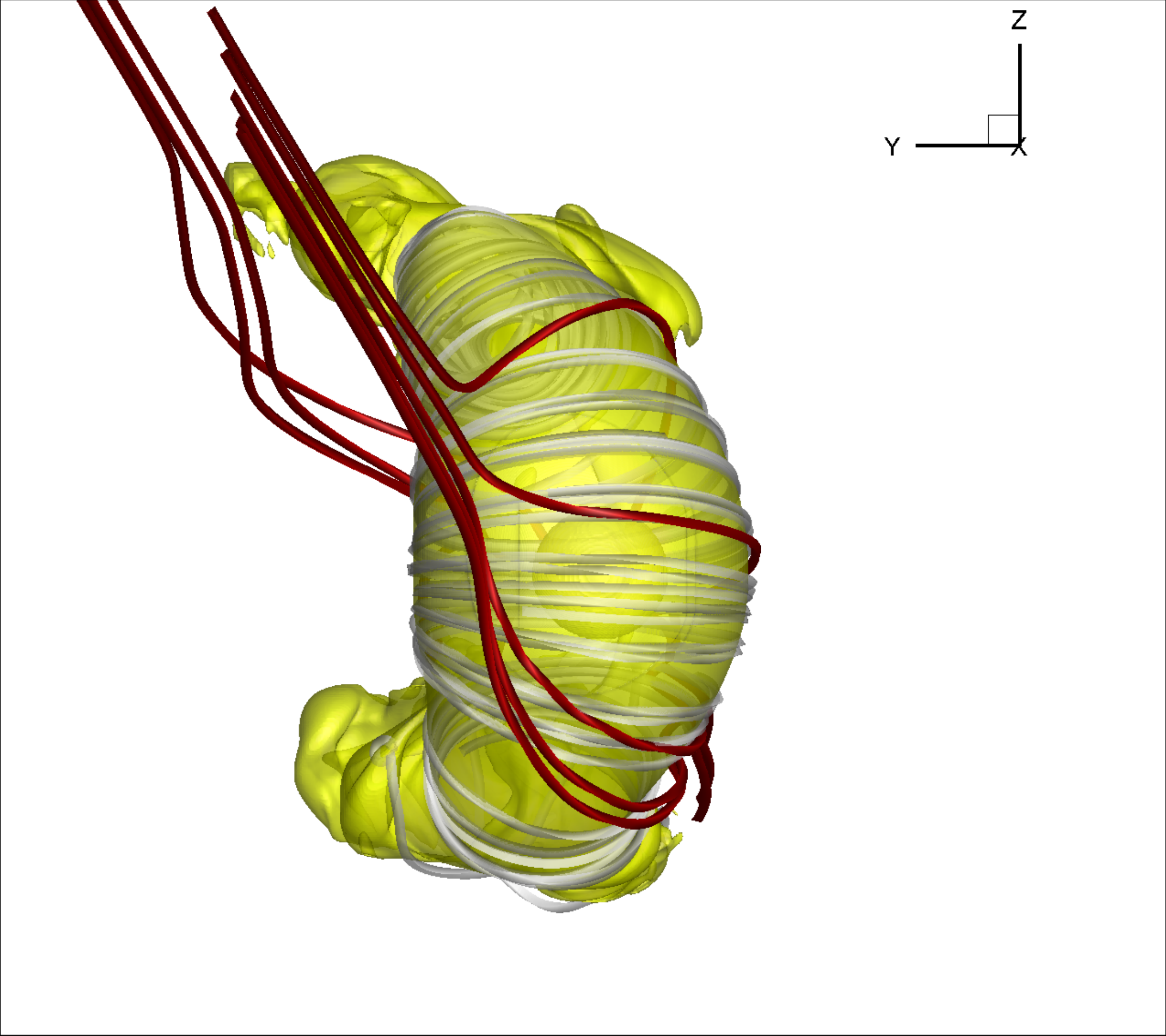}  
\caption{Two-lobe structure heliosphere for the case with interstellar magnetic field. 
The heliopause is captured at the iso-surface of $lnT=12.7$; the gray lines are the solar magnetic field lines; the red lines are the interstellar magnetic field. (a) Side View; (b) Nose view. }
\label{figure2}
\end{figure}

The lobes are turbulent both with and without the $B_{ISM}$ but the turbulence is stronger with a $B_{ISM}$. This can be seen in Figure 3a, 3b where the large-scale voids indicate that mixing of the ISM and solar material has taken place. This can be seen as well in the time evolution of the lobes where the large-scale voids convect (Figure 4). The presence of turbulent lobes has significant implications for reconnection and particle acceleration. Although once the solar wind and interstellar wind mix, the overall temperature drops, there are turbulent fronts where the temperature increases (Figure 4d). Again for this case, the magnetic and plasma pressure of the ISM in the region between the two lobes balances of the solar wind plasma within the lobes. 

\begin{figure}[htbp]
\centering
\includegraphics[width=0.4\textwidth]{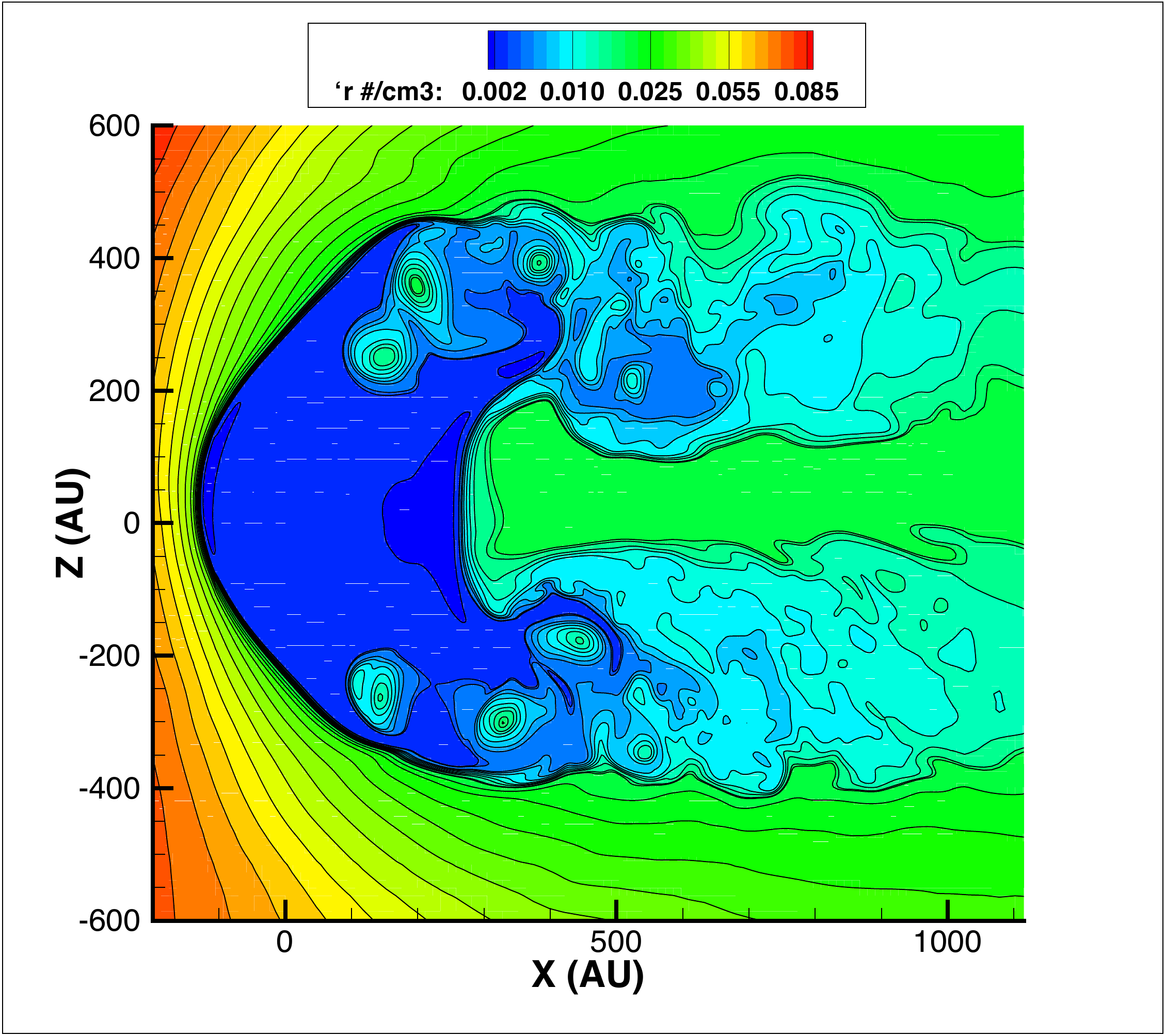}          
\includegraphics[width=0.4\textwidth]{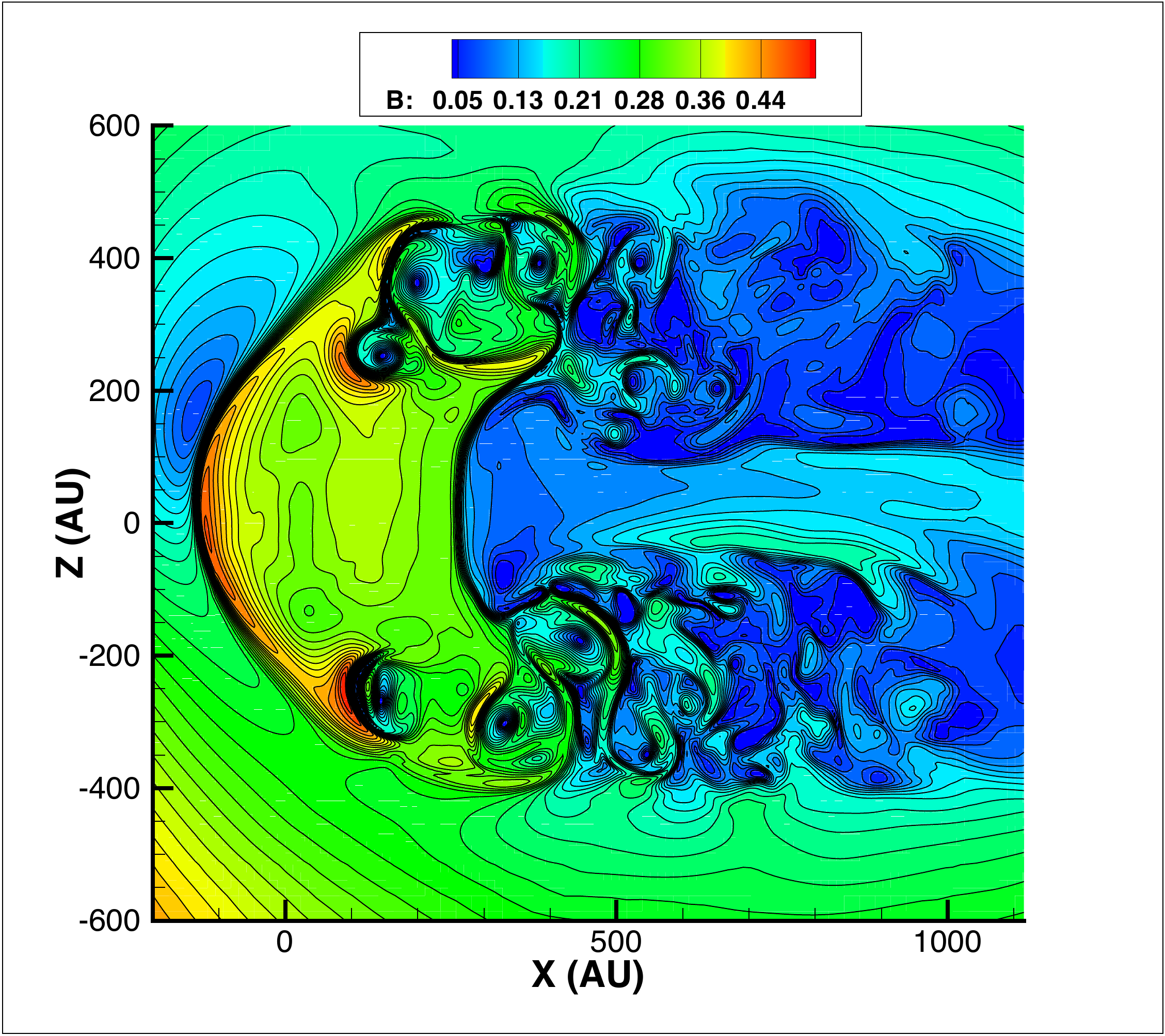}  
\includegraphics[width=0.4\textwidth]{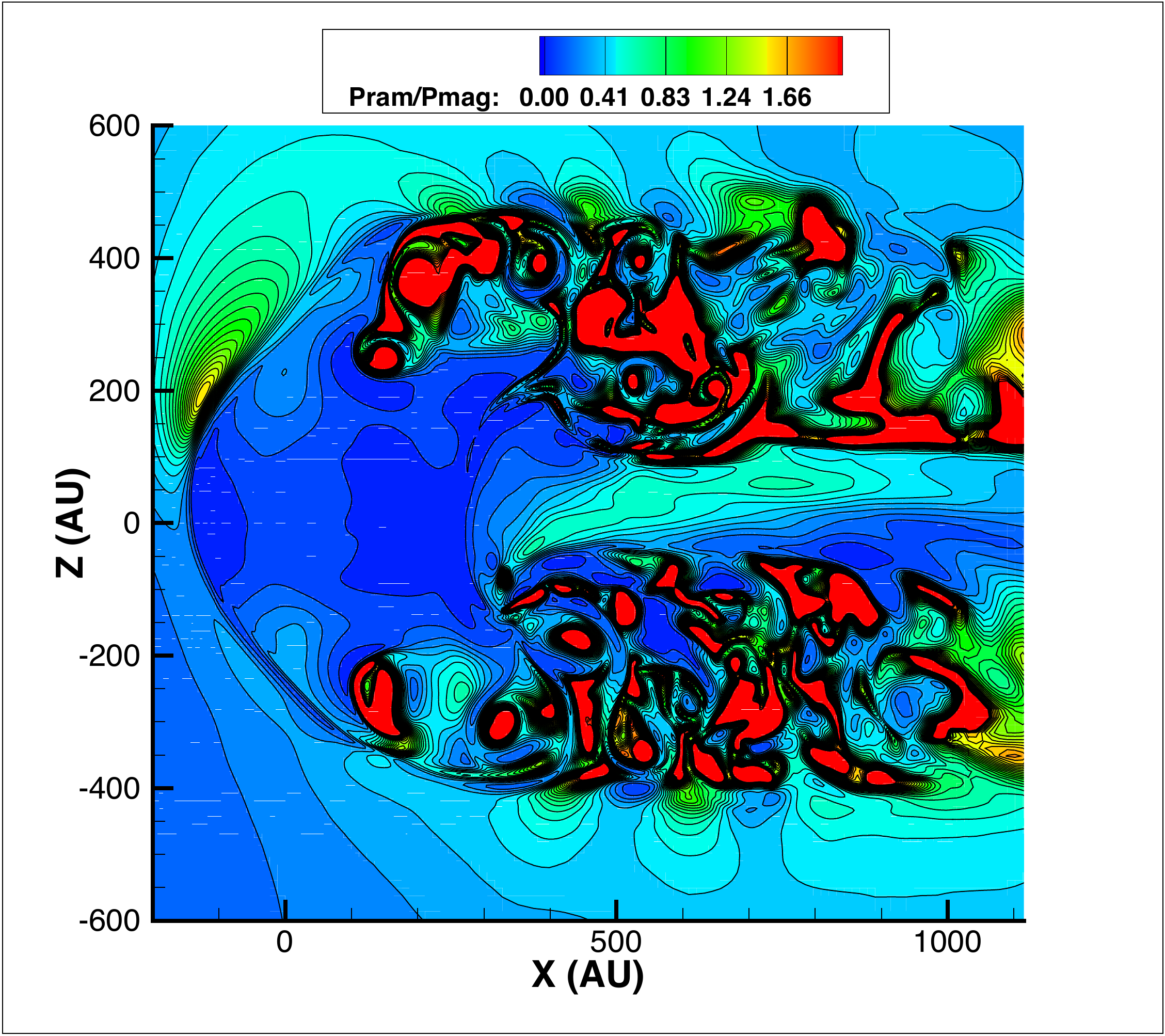}          
\includegraphics[width=0.4\textwidth]{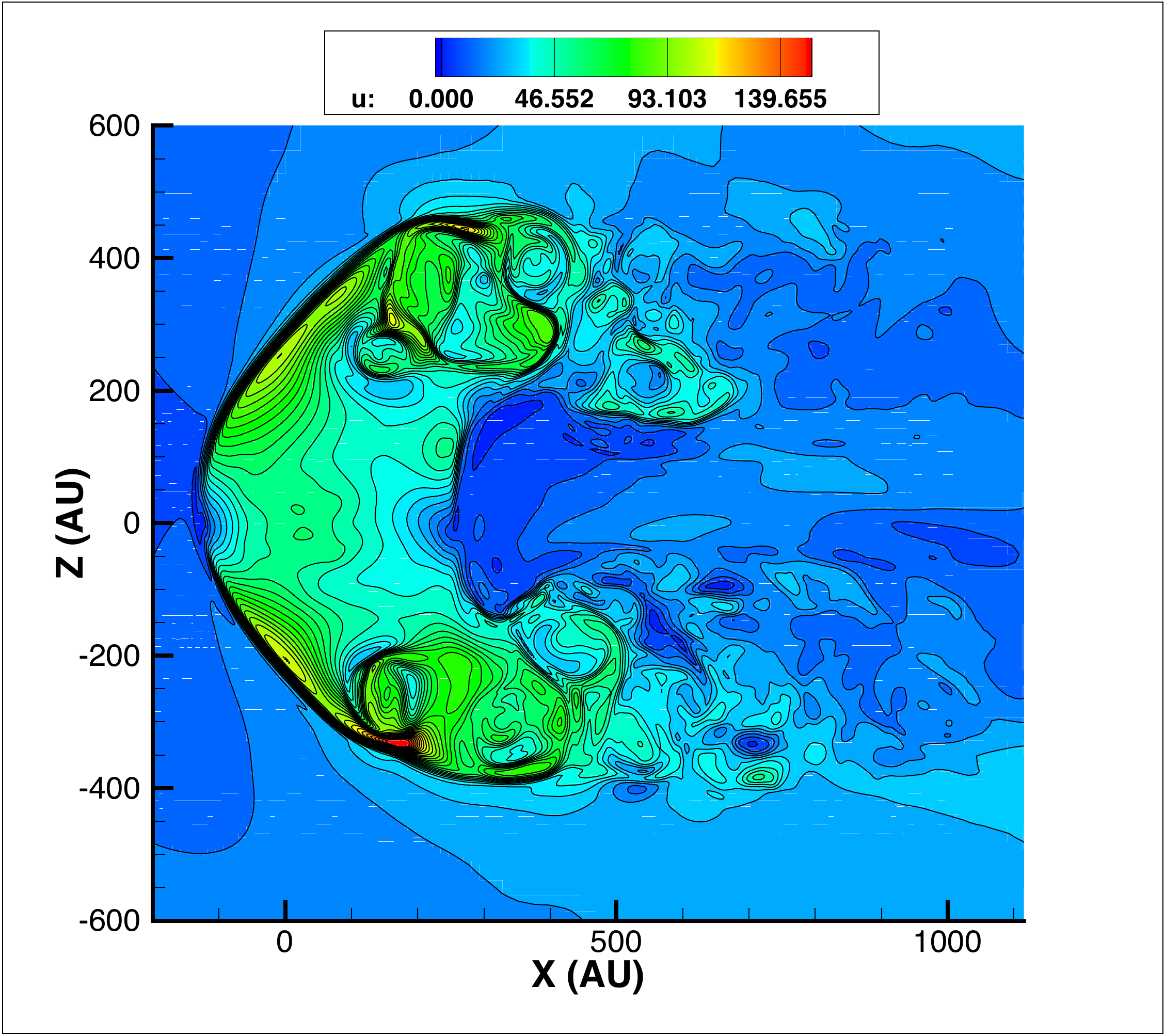}  
\includegraphics[width=0.4\textwidth]{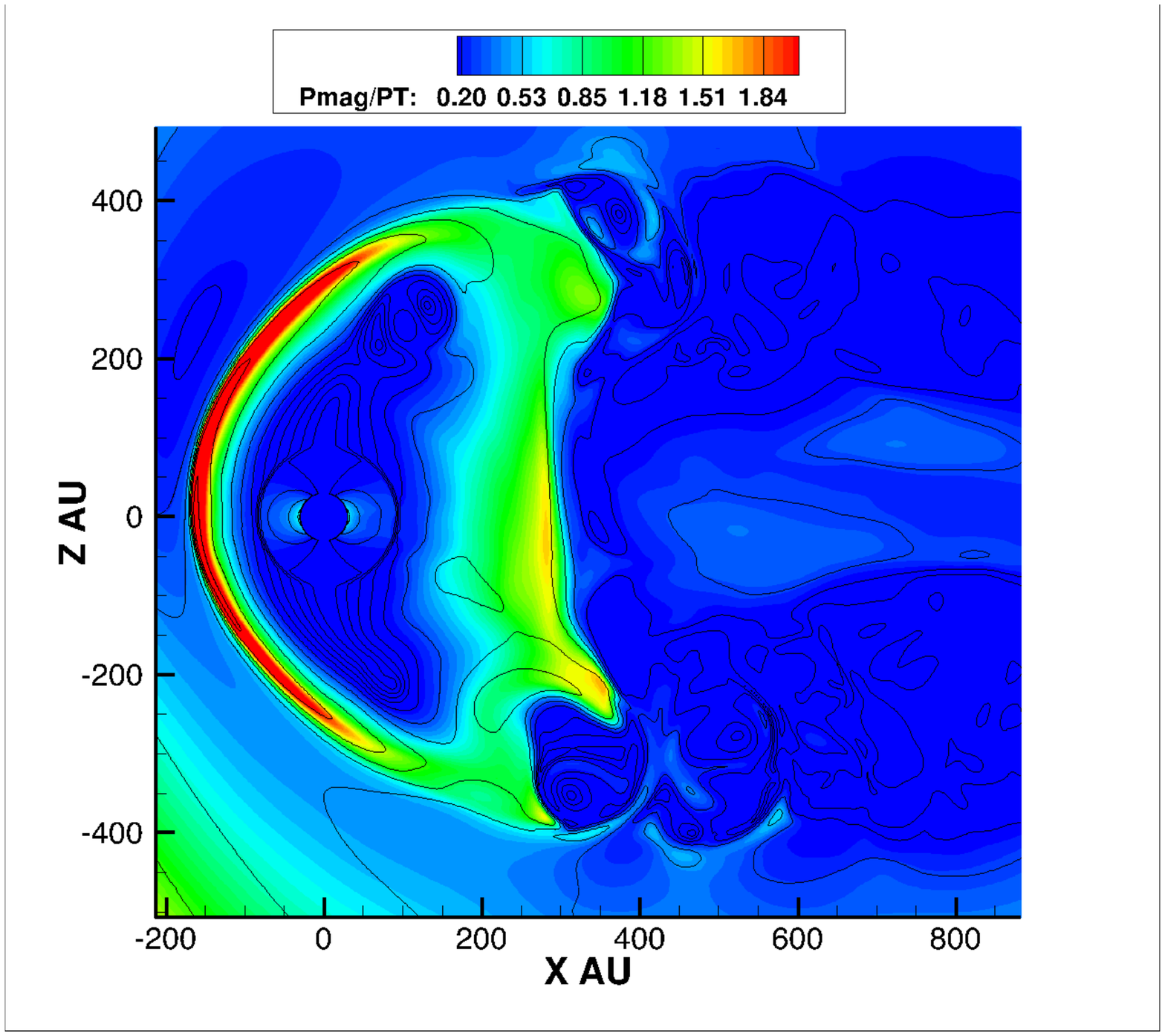}
\caption{Two-lobe structure - case with interstellar magnetic field. Cut at $y=150~AU$ (west flank as seen if the heliosphere is viewed from the ISM towards the nose of the heliosphere) at the end of the run at 659 years. Contours of: (a) density; (b) magnetic field; (c) Pram/Pmag; (d) speed. (e) Pmag/Pt; where Pt is the thermal pressure - this cut was done in $y=0~AU$. Line contorus are the total speed. }
\label{figure3}
\end{figure}

\begin{figure}[htbp]
\centering
\includegraphics[width=0.45\textwidth]{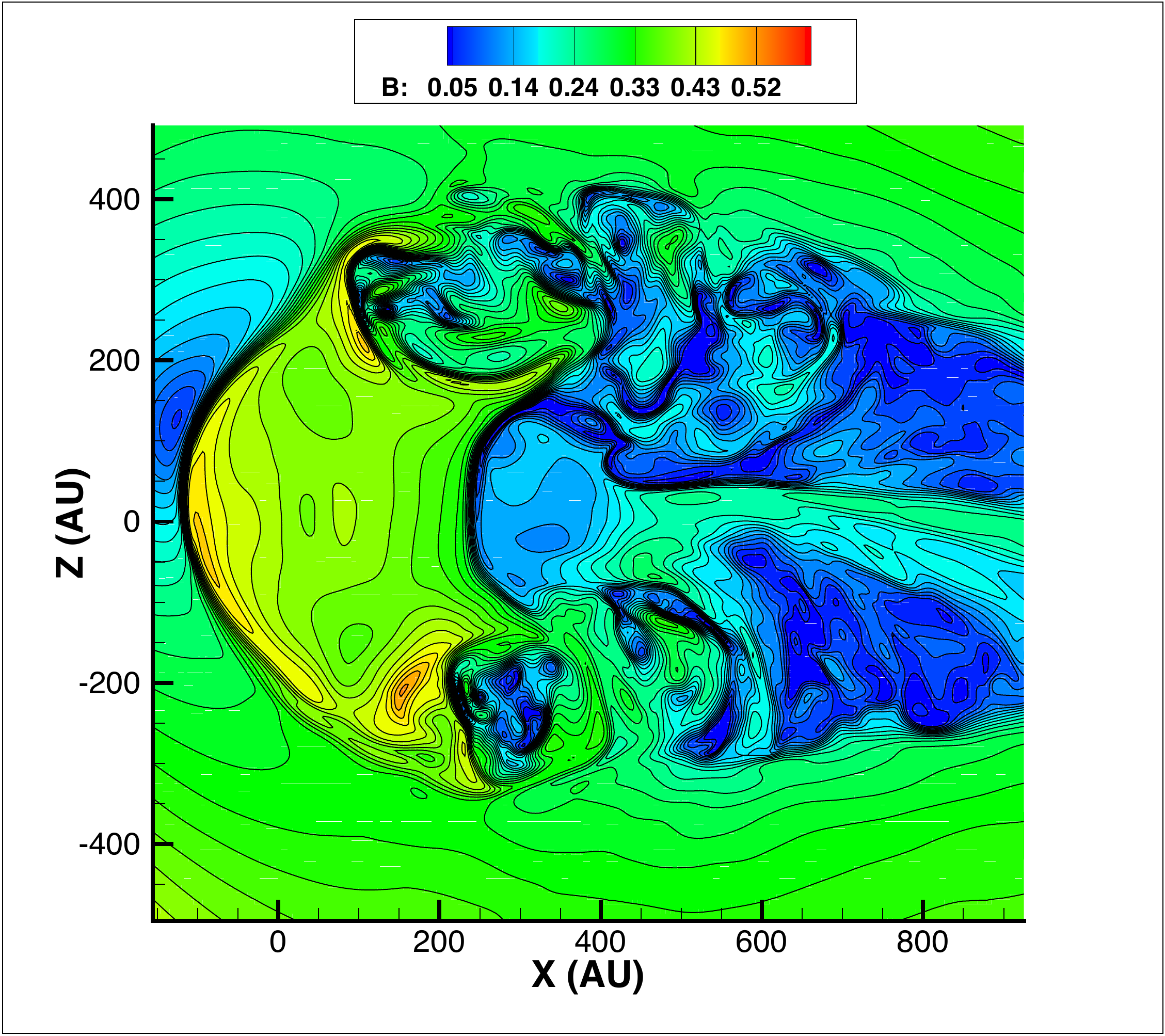}          
\includegraphics[width=0.45\textwidth]{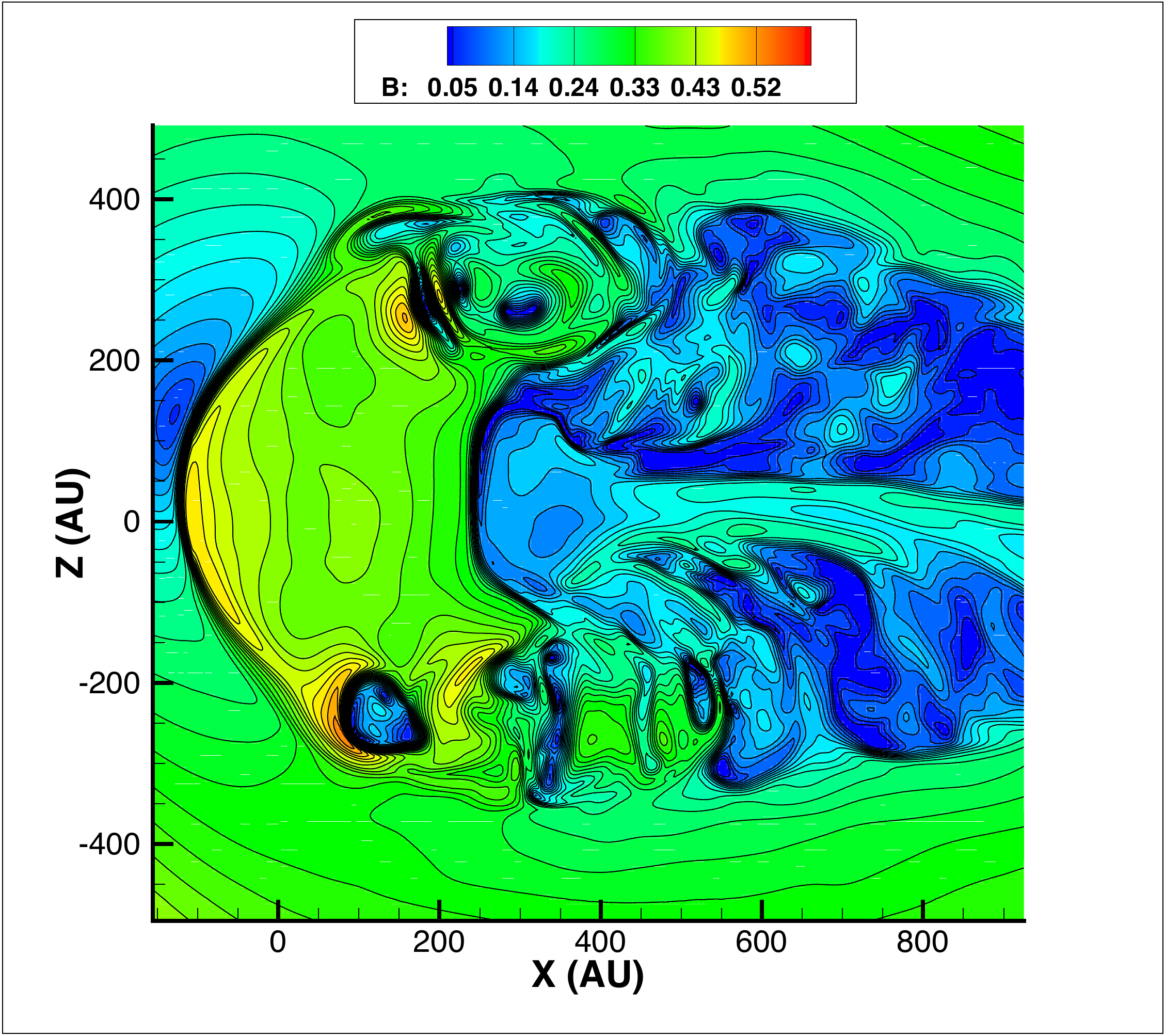}  
\includegraphics[width=0.45\textwidth]{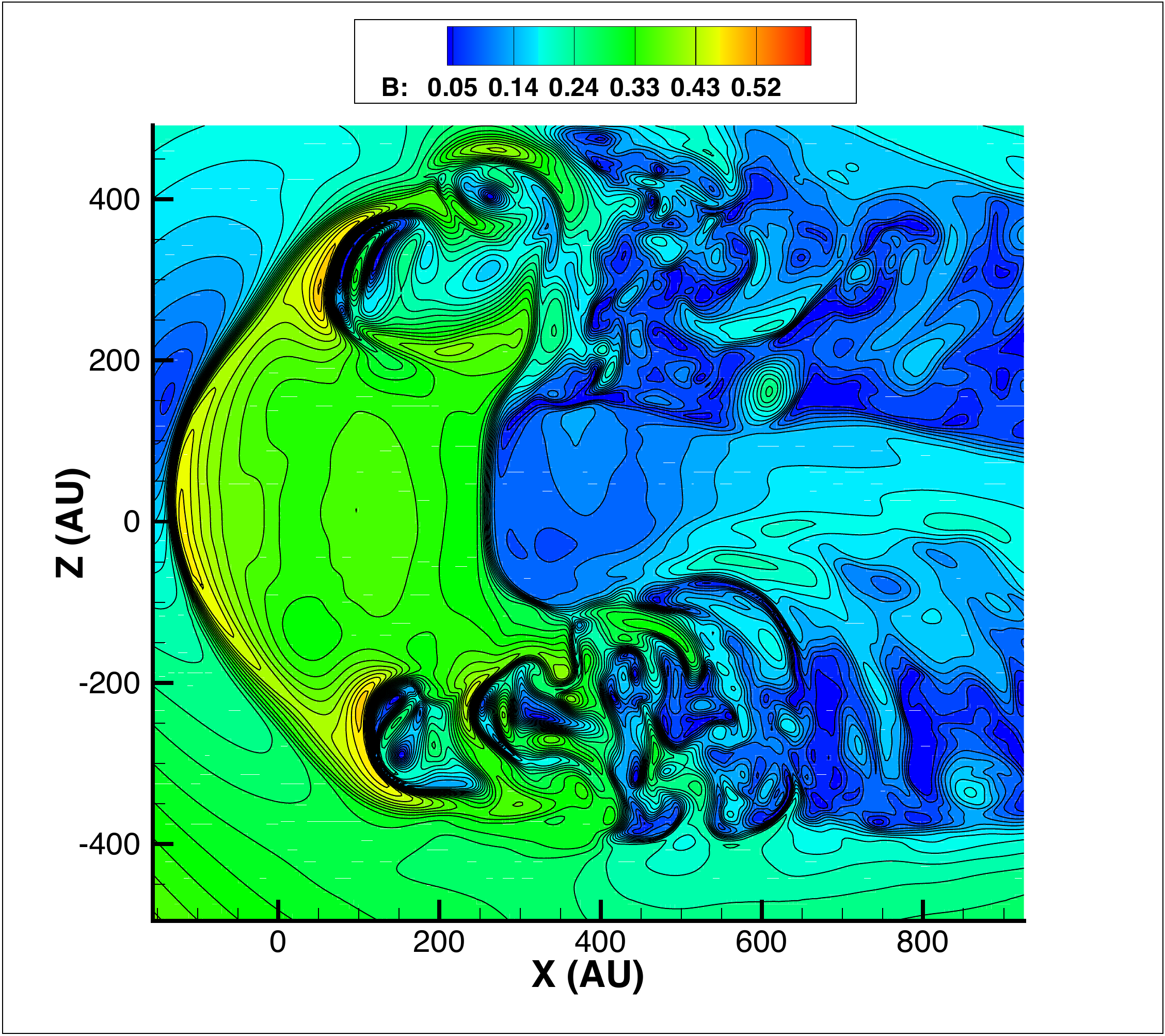}
\includegraphics[width=0.45\textwidth]{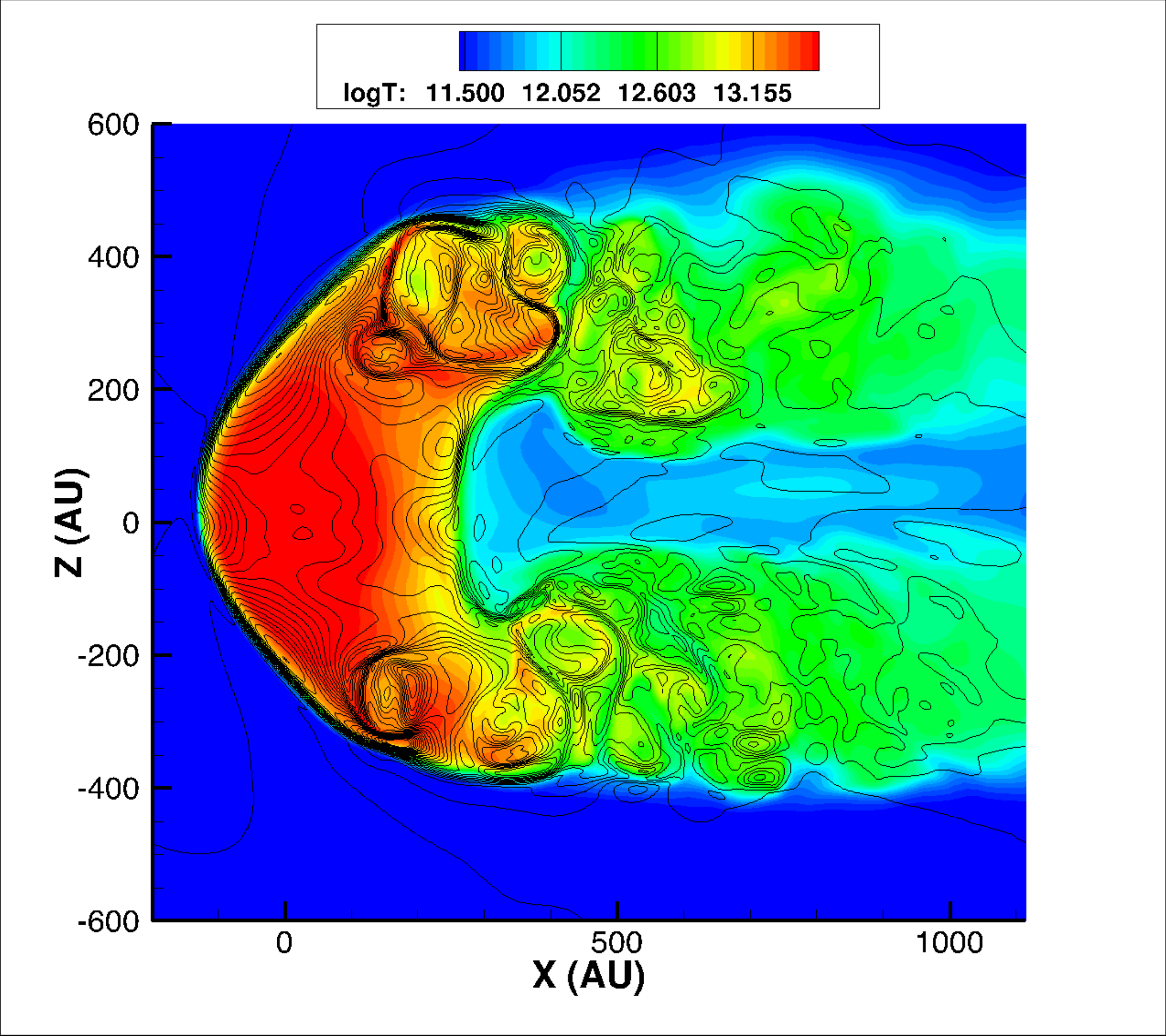}                    
\caption{Turbulent Flanks - case with interstellar magnetic field. Time Evolution at $y=150~AU$ (west flank). Panels (a-d) show contours of magnetic field (a) 375 years; (b) 404 years; (c) 546 years. The nature of the turbulent lobes can be seen. Panel (d) shows the contours colors of ln T with contour lines as the magnitude of the speed.}
\label{figure4}
\end{figure}

\begin{figure}[htbp]
\centering
\includegraphics[width=0.3\textwidth]{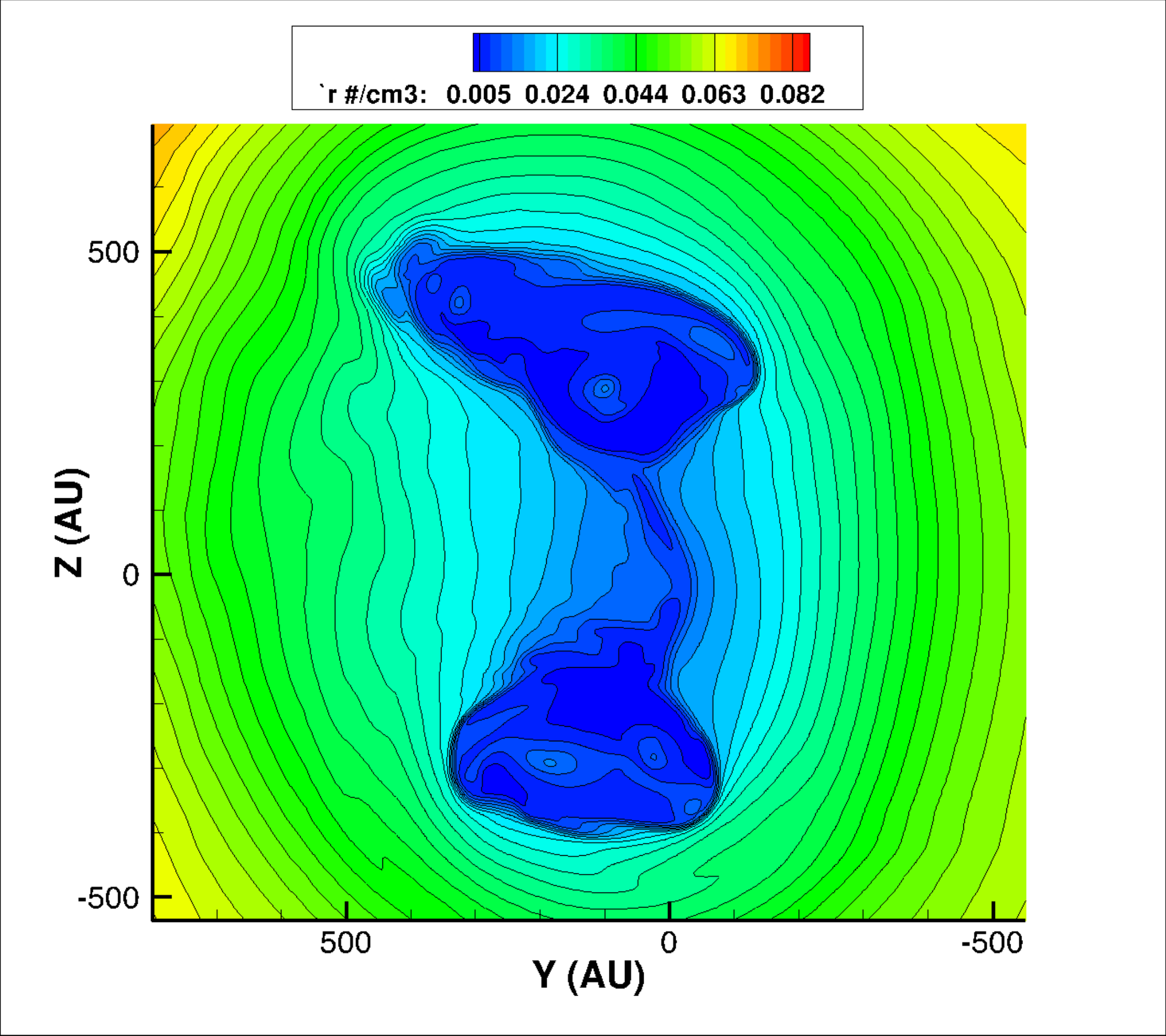}          
\includegraphics[width=0.3\textwidth]{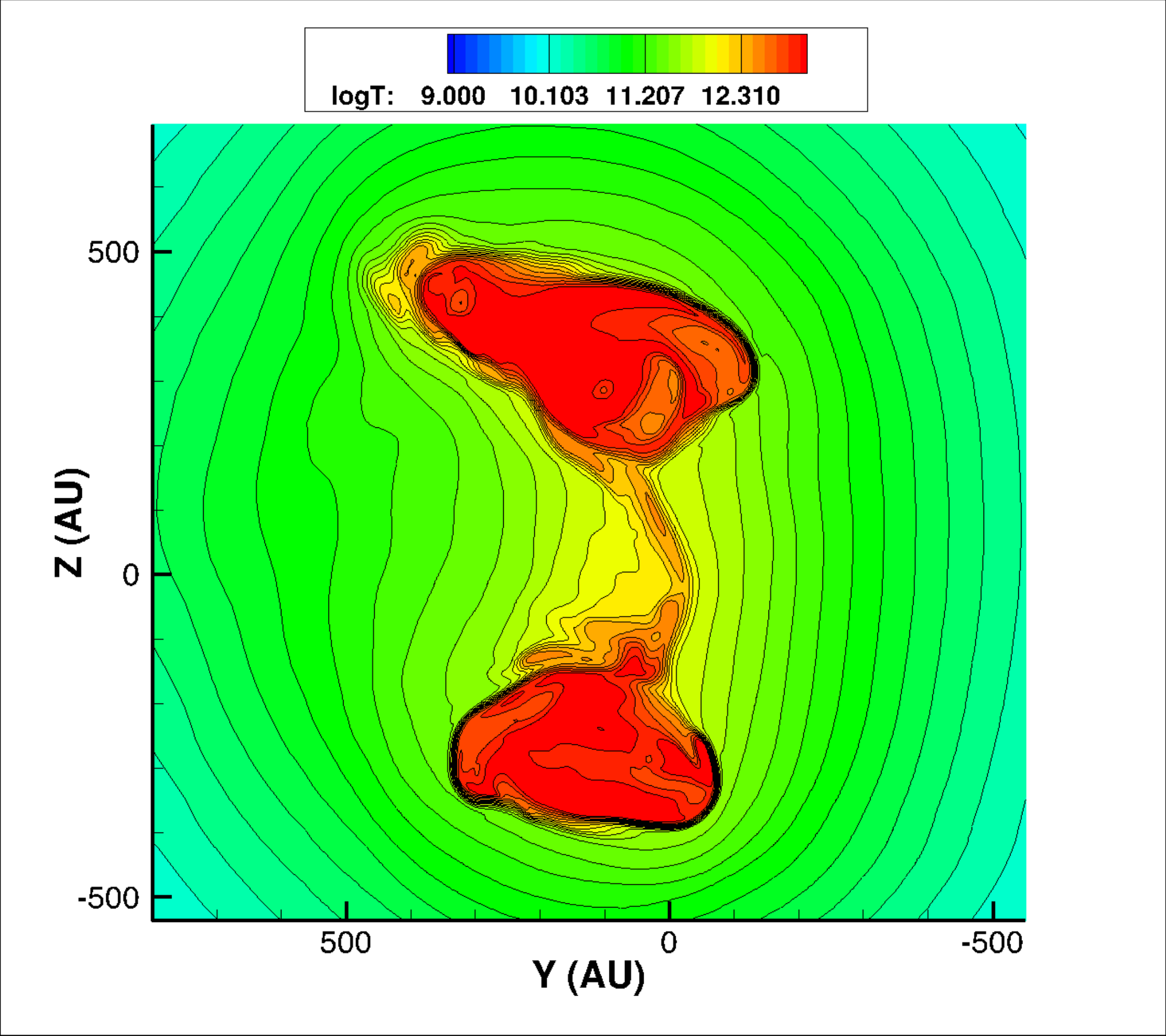}  
\includegraphics[width=0.35\textwidth]{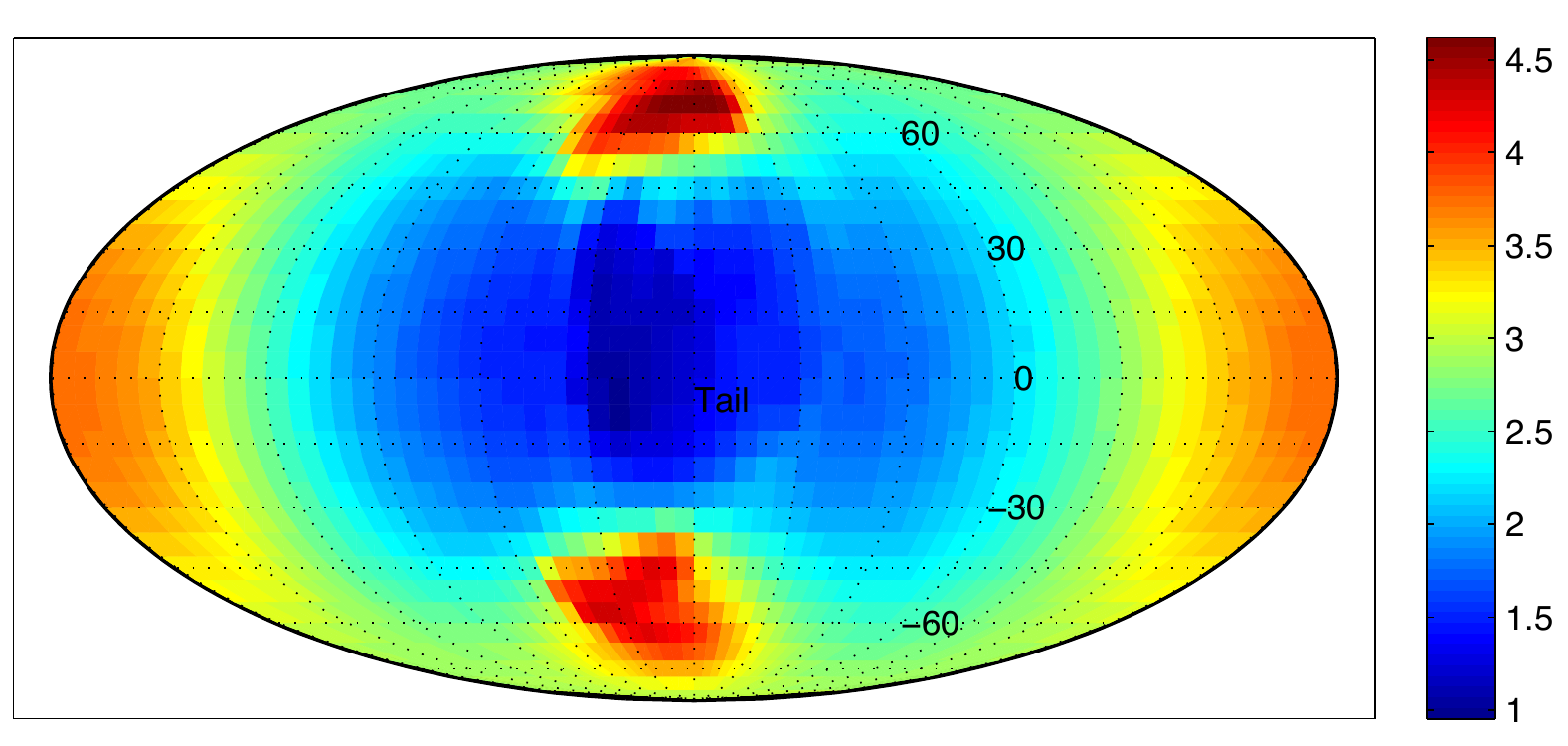}          
\caption{Two-lobe structure in density and temperature. Cut down the tail at $x=350~AU$ of the (a) density and (b) ln(T) as viewed from nose towards the tail. In (c) a proxy for the production of ENA (a line of sight integration of ion pressure times the neutral density from 100 to 350AU) viewed towards the tail in a Mollweide projection.}
\label{figure5}
\end{figure}

For astrophysical jets such as in Active Galactic Nuclei, the formation of a comet-like tail can occur, depending upon local conditions such as the ram pressure in the jet versus that of the ambient medium as well as the strength of instabilities that cause jet spreading (Morsony et al. 2013). 

The obvious question is why previous global MHD models did not produce the two-lobe structure. Earlier models either didn't include the solar magnetic field (Alexashov et al. 2004, Ratkiewicz \& Webb 2002) or included it as a dipole field (Opher et al. 2009, Pogorelov et al. 2007). Simulations with a dipole magnetic field or a varying tilt of the solar magnetic field were dominated by grid-induced reconnection around the solar equatorial plane that artificially eroded the solar magnetic field (Washimi et al. 2011, Provornikova et al. 2014, Pogorelov et al. 2013, Opher et al. 2011). The erosion of the magnetic field suppresses the confinement of the solar plasma and prevents the formation of the two-lobe structure. 

In the simulations presented here we used a monopole solar magnetic field. In reality because of the tilt of the solar magnetic field with respect to the rotation axis of the Sun there is a sector zone where the Parker azimuthal magnetic field periodically reverses direction. As long as in the solar magnetic field is not too strongly eroded by reconnection in the sector zone, the solar magnetic field will remain strong enough to confine the solar wind plasma and the two-lobe structure of the heliosphere will survive. The Voyager observations indicate that while some erosion of magnetic flux has taken place the solar magnetic field still survives (Richardson et al. 2013). However we do expect that reconnection in the sector zone and with the interstellar magnetic field will lead to some erosion of the two lobes (Opher et al. 2011, Drake et al. 2010, Swisdak et al. 2013, Strumik et al. 2013). Future tail ENA observations by IBEX or CASSINI might be able to determine the time evolution of the two-lobe structure. 

The overall two-lobe structure is consistent with the ENA images from IBEX that for the first time mapped the heliotail. Such images show two lobes (McComas et al. 2013) with an excess of low energy ENA ($<1keV$) and a deficit at higher energy ($>2keV$) around the solar equator. A view of the data from the MHD simulation  (Figure 5) also reveals two lobes centered at high latitude (seen in the lower density and higher temperature of the heliosheath). It is possible that the denser ISM between the lobes at low-latitude is responsible for the two lobes seen in maps of the low energy IBEX data (via a pick-up ion contribution in the ISM or a secondary ENA process). Another possibility is that the turbulent lobes accelerate particles that can produce such signatures. In Fig. 5(c) we show a proxy of the high energy ENA that are expected to come from the heliosheath. Two regions of strong emission in the North and South are visible. This image can be compared to the high-energy ENA maps from IBEX (McComas et al. 2013) where similar lobes are also visible. A more detailed exploration of potential ENA signatures of the two-lobe heliosphere is required but is beyond the scope of this paper. 

The ENA images from Cassini (Krimigis et al. 2009, Dialynas et al. 2013) (at a much higher energies $5-55~keV$) revealed intensities that were comparable in the direction of the nose and tail. The observers therefore concluded that the heliosphere might be ``tailless'' because the emission from these high energy ENAs is believed to come from the heliosheath. The two-lobe heliosphere is in fact almost ``tailless'' with the distance down the tail to the ISM between the lobes being nearly equal to the distance towards the nose. If in fact the two-lobes are as eroded as we found in the case when $B_{ISM}$ is present, there will be a much smaller contribution to high energy ENA coming from the tail than expected based on a comet-like heliosphere.

\section{Discussion}

Other magnetospheres (such as Earth (Siscoe et al. 2004) and Saturn (Jia et al. 2012, Zieger et al. 2010)) exhibit a two-lobe structure. These structures are not related to the phenomena discussed in this paper – but to reconnection of the downtail component of the draped solar magnetic field that produces a dominant midtail x-line. The key ingredient here is the solar magnetic field that confines and collimates the solar wind and the ISM pressure that maintains the separation of the two lobes in the tail. 

We expect that such a two-lobe structure might be present in other outflows (astrospheres and exo-solar bubbles) where the magnetic field is present. For different interstellar conditions the lobe behavior will be different. For example, with a stronger wind the two lobes will be pushed closer together. 

The two-lobes structure of the heliosphere is similar to astrophysical jets in protostellar systems (Fendt \& Zinnercker 1998, Gueth \& Guilloteau 1999) and clusters of galaxies (Owen \& Rudnick 1976, Douglas et al. 2011) which are collimated by the magnetic field and are often seen as bent as a result of interaction of the jets and their surrounding media.

\acknowledgments
The authors would like to thank the anonymous referee for very helpful comments. They would also like to acknowledge helpful discussions and comments from Drs. Jonathan McKinney, Alan Marscher, Elizabeth Blanton and Randy Jokipii. The authors would like to thank the staff at NASA Ames Research Center for the use of the Pleiades supercomputer under the award SMD-14-4986. M. O. and B. Z. acknowledge the support of NASA Grand Challenge NNX14AIB0G and NASA award NNX13AE04G. J. F. D acknowledges the support of NASA Grand Challenge NNX14AIB0G and NASA award NNX14AF42G. In memory of my mother E. Opher.

\end{document}